# Contrast-enhanced dual-energy subtraction imaging using electronic spectrum-splitting and multi-prism x-ray lenses

Erik Fredenberg,[a] Björn Cederström,[a] Mats Lundqvist,[b] Carolina Ribbing,[c] Magnus Åslund,[b] Felix Diekmann,[d] Robert Nishikawa,[e] and Mats Danielsson[a]

[a]Department of Physics, Royal Institute of Technology (KTH), AlbaNova University Center, 106 91 Stockholm, Sweden;

[b]Sectra Mamea AB, Smidesvägen 5, 171 41 Solna, Sweden;

[c]The Ångström Laboratory, Uppsala University, 751 21 Uppsala, Sweden;

[d]Department of Radiology, Charité - University hospital, Charitéplatz 1, 101 17 Berlin, Germany;

[e] Deptartment of Radiology, University of Chicago, Chicago, IL, 60510, USA;

## ABSTRACT

Dual-energy subtraction imaging (DES) is a method to improve the detectability of contrast agents over a lumpy background. Two images, acquired at x-ray energies above and below an absorption edge of the agent material, are logarithmically subtracted, resulting in suppression of the signal from the tissue background and a relative enhancement of the signal from the agent. Although promising, DES is still not widely used in clinical practice. One reason may be the need for two distinctly separated x-ray spectra that are still close to the absorption edge, realized through dual exposures which may introduce motion unsharpness.

In this study, electronic spectrum-splitting with a silicon-strip detector is theoretically and experimentally investigated for a mammography model with iodinated contrast agent. Comparisons are made to absorption imaging and a near-ideal detector using a signal-to-noise ratio that includes both statistical and structural noise. Similar to previous studies, heavy absorption filtration was needed to narrow the spectra at the expense of a large reduction in x-ray flux. Therefore, potential improvements using a chromatic multi-prism x-ray lens (MPL) for filtering were evaluated theoretically. The MPL offers a narrow tunable spectrum, and we show that the image quality can be improved compared to conventional filtering methods.

## 1. INTRODUCTION

Contrast agents are used within many fields of medical x-ray imaging to improve the contrast between structures of similar density and atomic number. In particular, tumors are enhanced since the angiogenesis associated with the growth of lesions leads to an increased permeability and retention of the agent. The visibility of breast tumors can for instance be improved in computed tomography by intravenous administration of iodinated contrast agent.[1] For standard screen-film or even digital mammography, however, the relatively low contrast resolution limits the detectability even of contrast-enhanced lesions.[2]

To emphasize the agent above what is possible with plain absorption imaging, contrast-enhanced dual-energy subtraction imaging, here referred to as dual-energy subtraction (DES) imaging, has been proposed and investigated for mammography[2–5] and other x-ray imaging modalities.[6–8] It relies on the fact that, at a material specific energy, the absorption coefficient of the contrast agent changes rapidly. A so called absorption edge is caused by the radically increased cross section for photoelectric interaction between an incident photon and an

---

Electronic mail: fberg@mi.physics.kth.se

atom in the material as the photon energy reaches the binding energy of an atomic shell. Iodine, for instance, has a K absorption edge at 33.2 keV. To perform DES imaging, two images are acquired simultaneously or consecutively with two different energy spectra; one with a mean energy below and one with a mean energy above an absorption edge of the contrast agent. By combining the two images in the logarithmic domain, the signal from any pair of materials in the object can be made to cancel, for instance glandular and adipose tissue in mammography. Constituents with a different attenuation shift between the high and low energy images are still visible, and the signal from the contrast agent can hence be enhanced over a cluttered background.

To efficiently perform DES imaging it is important to use two narrow spectra straddling the edge. One way of providing the two spectra is to use two different anode materials with appropriate absorption filtering.[2,4] Drawbacks of this dual spectra (DS) approach include the need for two separate exposures, and a limited effectiveness due to the large spread of the spectra and low flexibility in the choice of anode materials. These are two reasons to why DES, although an old idea, has not yet evolved into routine clinical practise.

Another option to provide the two spectra is to center a single absorption filtered spectrum on the absorption edge and use an energy sensitive detector, for instance a sandwich detector[9] or two different detector materials,[10] to split the spectrum electronically at the edge. One recently developed detector that has been briefly investigated for DES imaging is a photon-counting silicon-strip detector with two energy thresholds.[5] It is similar to the detector in an existing full-field digital mammography system.[11] The performance of electronic spectrum splitting (ES) depends on how much the spectrum can be narrowed down around the edge. When absorption filtering is used there is thus a trade-off between image quality and efficiency since heavy absorption filtering and a low acceleration voltage result in poor photon economy.

The multi-prism lens (MPL) is a refractive x-ray lens with chromatic properties, which has been shown to work well at x-ray energies relevant to mammography.[12] It can be employed as an energy filter, discriminating against high as well as low energy photons, thus providing a narrow spectrum with better photon economy than is achievable with absorption filtering.[13] MPL filters can potentially improve on both the DS and ES approaches for DES imaging. Narrow spectra can be produced at arbitrary energies from a wide tungsten spectrum, thus reducing the energy spread and eliminating the need for two anode materials in the DS method. Alternatively, ES of a single MPL-filtered spectrum tuned to center around the energy of the absorption edge provides more optimal conditions for subtraction imaging than is practically achievable with absorption filtering. In addition, the tunability of the MPL filter opens up the possibility to use contrast agents with absorption edges closer to the ideal imaging energies of the imaged object, thus providing a regular absorption image of high quality in addition to the DES image. Zirconium with a K-edge at 18.0 keV has for instance been suggested for mammography.[14]

In this study, ES will be further investigated for the aforementioned silicon-strip detector in a mammography model with iodinated contrast agent. The improvement over absorption imaging, and the efficiency compared to a near-perfect detector is evaluated experimentally. A theoretical model including detector imperfections is developed and verified by measurements. By using this model, the potential improvement of ES by MPL filters is evaluated. Investigations of MPL filters for the DS approach and the possibility to use novel contrast agents are left for future studies.

## 2. MATERIAL AND METHODS

### 2.1. Dual energy subtraction

Two different methods for calculating DES signals appear frequently in the literature; the weighted logarithmic subtraction method,[2,5,10] and the material basis plane decomposition method.[3,4,6] In this study, the former approach will be used.

The number of counts in a detector, assumed ideal, from an energy interval $\Omega$ of an impinging spectrum $I_0(E) = N_0 \Phi(E)$ after passing a breast of thickness $d$ and linear attenuation coefficient $\mu(E)$ is

$$n = N_0 \int_\Omega \Phi(E) \cdot \exp(-\mu(E)d) \cdot dE, \qquad (1)$$

where $N_0$ is the number of impinging photons and $\Phi(E)$ is the spectral distribution with sum unity. Then, using indices $hi$ and $lo$ to denote high and low energy bins, respectively, the subtracted intensity of the DES image is calculated with the weighted logarithmic subtraction method according to

$$
\begin{aligned}
I_\text{S} &= \ln n_\text{hi} - w \ln n_\text{lo} \\
&= \ln N_{0,\text{hi}} - w \ln N_{0,\text{lo}} + \ln \int_\Omega \Phi_\text{hi}(E) \cdot \exp(-\mu(E)d) \cdot dE - w \ln \int_\Omega \Phi_\text{lo}(E) \cdot \exp(-\mu(E)d) \cdot dE,
\end{aligned}
\quad (2)
$$

where $w$ is a weight factor. Since the first two terms are constant, they will cancel when forming the DES signal difference between two areas in the breast with different attenuation,

$$\Delta S_\text{S} = \left| I_\text{S}^1 - I_\text{S}^2 \right|, \quad (3)$$

and hence the signal difference is independent of exposure.

In this study, $w$ was chosen so as to minimize the noise from the anatomical background, in accordance with previous studies.[5] Although disregarding the statistical noise and the iodine signal, this procedure will generally result in the optimal image quality, as is further discussed in Section 2.4. The anatomical noise was calculated as the standard deviation of $I_\text{S}$ measured over a range of glandular fractions, $g_i$;

$$\sigma_\text{w}(w) = \left( \frac{1}{m-1} \sum_i^m \left[ I_\text{S}(g_i, w) - \overline{I_S(w)} \right]^2 \right)^{1/2}. \quad (4)$$

Note that this quantity is also exposure-independent. For the optimization, $w$ was chosen so that $\sigma_\text{w}$ is minimized when using a linear function $g_i$ from 0.1 to 0.9.

**2.2. Experimental evaluation**

DES measurements with ES were performed in a set-up similar to a scanned-slit mammography geometry (Fig. 1). The source is a $0.4 \times 12$ mm tungsten target x-ray tube[*] with variable acceleration voltage 10-60 kVp. It is viewed at a 2.2° anode angle, which yields an effective source size of approximately $400 \times 461$ $\mu$m. The x-ray beam was filtered with aluminum absorption filters and collimated with a slit before passing the object and reaching the detector.

Two different detector assemblies were used for the measurements; a cadmium-zinc-telluride (CZT) compound solid-state detector[†] with approximately 1 keV energy resolution to simulate nearly perfect spectrum splitting, and a silicon-strip detector[11] with two energy thresholds to simulate more realistic conditions for medical imaging. The one pixel CZT detector was coupled to a 500 channel multi-channel analyzer, which was read out by a computer. In that way, the energy threshold for splitting the spectrum can be set with good resolution and adjusted after the experiment.

The silicon strip detector is 128 pixels wide with a 50 $\mu$m strip-pitch in an edge-on arrangement. It is wire-bonded to a 128-channel pulse-counting application specific integrated circuit (ASIC) equipped with two adjustable energy thresholds for each channel. To reject noise, the low-energy threshold was set to discriminate against pulses below approximately 5 keV. The high-energy threshold was set to approximately 33 keV, corresponding to the K-edge of iodine. The detector assembly is thus photon-counting with virtually no electronic noise present, and all detected photons are divided into two bins according to their energy.

Between the collimator slit and the detector, a motorized stage was placed to scan an object across the beam and thus acquire an image. The phantom that was used for the measurements is made up of three parts (Fig. 1). A 10 mm thick polymethyl methacrylate (PMMA) slab has containers 1-9 mm deep which were filled with iodinated contrast agent.[‡] The iodine concentration was approximately 3 mg/ml, which has been found to be a realistic concentration for tumor uptake.[1] In front of the PMMA slab is placed a 35 mm thick PMMA box which can be filled with PMMA and olive oil in different compositions to simulate breast tissue. PMMA cylinders with diameters 5-10 mm immersed in oil were used to simulate anatomical clutter, and a 1.75 cm thick PMMA slab in oil was used to simulate a homogenous breast with 50% glandularity. PMMA and olive oil were chosen because the difference in linear attenuation is close to the difference for fibroglandular and adipose breast tissue.[5] There is also a wedge phantom composed of PMMA and oil with PMMA fractions ranging from 10% to 90% to simulate a variety of glandularities. The average glandular dose (AGD) to the phantom was calculated by applying normalized glandular dose coefficients[15] to a measured spectrum.

---
[*]Philips PW2274/20 with high tension generator PW1830
[†]Amptek XR-100T-CZT
[‡]Ultravist 370, BayerSchering, Germany

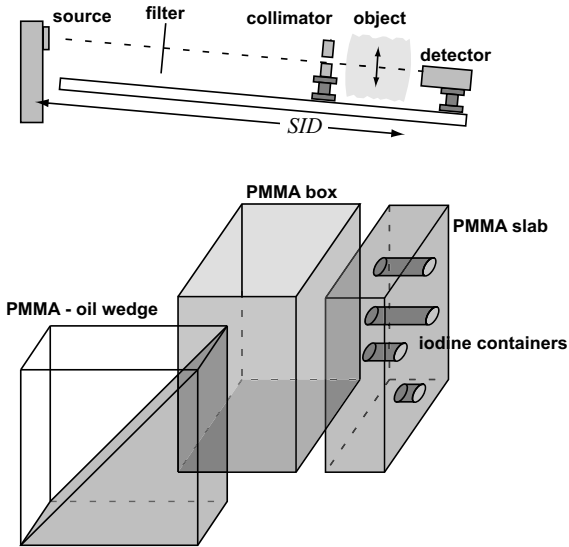
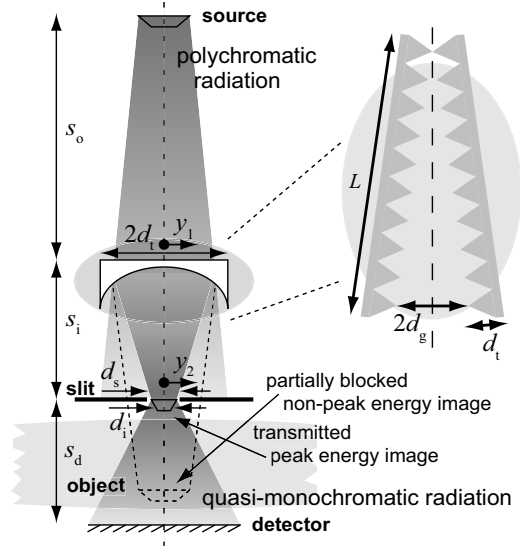

**Figure 1. Top:** The set-up used for the DES experiments, where the detector is either a CZT or a silicon-strip detector. **Bottom:** The phantoms; a 35 mm thick PMMA box filled with PMMA and olive oil to simulate breast tissue, a 1 cm PMMA slab with 9 iodine containers ranging from 9 to 1 mm (only 4 are shown for clarity), and a PMMA-to-oil wedge phantom to simulate a range of glandularities.

**Figure 2.** The theoretical MPL set-up with $s_o = 759$ mm and $s_i = 310$ mm. A $d_s = 10$ $\mu$m slit in the image plane of the lens blocks radiation that is out of focus before passing the object. The lens has tooth height $d_t = 100$ $\mu$m and focal length 220 mm.

### 2.3. Theoretical model

To verify the experimental results and to make further predictions, the experiment was modeled using the MATLAB software package[§] with published linear absorption coefficients[16] as input.

Several detector effects have to be taken into account. As for the CZT detector, it suffers from hole-tailing due to trapped charges in crystal imperfections, and from a limited energy resolution due to the small amount of electron-hole pairs that are released at each photon interaction event. Although hole-tailing extends only in the negative energy direction (charges are lost) and both of the effects might vary with energy, for simplicity they were joined into a gaussian that was convolved with the high and low energy spectra. The quantum efficiency of the 2 mm thick CZT crystal with the 0.3 mm beryllium window in front was also taken into account.

The silicon strip detector does not suffer from hole-tailing to any large extent, but the limited energy resolution due to a low number of released charge pairs is present and was again modeled by a gaussian. Additionally, all thresholds are set separately for individual channels and therefore vary somewhat. This effect degrades the energy resolution further and was modeled by convolving the spectrum with another gaussian. The 500 $\mu$m thick silicon wafer was arranged at an angle of 4°, which yields an effective detector thickness of 7.2 mm, and the quantum efficiency could be calculated accordingly. No secondary interactions of scattered photons were considered. Charge sharing is a fourth effect that has to be taken into account for the position sensitive silicon-strip detector. If a photon interacts close to the midpoint between two strips, some charge might spill over to the adjacent strip and two photons of lower energy are detected instead of one high-energy photon. The ASIC that was used for the experiment is equipped with anti-coincidence logic which rejects the lowest pulse and puts the highest pulse in the high-energy bin regardless of pulse height. These effects have been investigated previously for a similar detector,[11] and the same calculations were applied here to model the effect.

---

[§]The MathWorks Inc., Natick, Massachusetts

Several of the model parameters mentioned above are unknown and the model was bench-marked to the experiments. The iodine concentration, first of all, may not be very exact, and was therefore measured in the absorption images. Secondly, the width of the energy resolution gaussian for both detectors was set so that the theoretical subtraction signal from the iodine containers matched the measurements with the CZT detector. The threshold resolution of the silicon detector, finally, was set so that the iodine signal in the acquired images matched the model.

## 2.4. Dual energy figure of merit

There is no standard method to quantify the image quality in DES, and we introduce a figure-of-merit based on the signal-difference-to-noise ratio (SDNR). This SDNR must take into account both the structural noise from the anatomical clutter as well as the statistical noise. It is important to realize that these have completely different frequency distributions. Also note that we are here interested in large targets (tumors with iodine uptake), $i.e.$ low spatial frequencies. Observer models based on integration of the signal template and NEQ(f) were not used, partly due to that the experimental data was too limited for reliable NPS calculations. Instead we choose a crude method in the spatial domain, where the noise is quantified by the standard deviation of the background. This, however, is not just done on a pixel basis. Instead the image is subsampled (binning without changing the image average) into ROIs of variable size. This ROI size ($M \times M$ pixels) corresponds to the size of the details in the image that we are interested in. The method is obviously related to the NEQ-integration, since the image binning corresponds to a sampling of the NPS, in which high spatial frequencies are suppressed.

We define our figure-of-merit through

$$\text{SDNR}_{tot,M} = \frac{|S_{target} - S_{bg}|}{\sigma_{tot,M}} = \frac{|S_{target} - S_{bg}|}{\sqrt{\sigma_{quant,M}^2 + \sigma_{clutter,M}^2}} \quad (5)$$

We will assume that the system is quantum limited. After the logarithmic subtraction the quantum noise is given by (since the noise is small compared to the signal and $\ln(1+x) \approx (1+x)$ for $x \ll 1$)

$$\sigma_{quant,1}^2 = 1/N_{hi} + w^2/N_{lo}. \quad (6)$$

For a system with white noise only we have

$$\sigma_{quant,M} = \sigma_{quant,1}/M, \quad (7)$$

whereas we expect a much smaller reduction of $\sigma_{clutter,M}$ with increasing $M$ due to the lumpiness of the structure.

The wedge-based definition of anatomical clutter noise, $\sigma_w$, used for optimization of $w$ is not a realistic one here, since it would overestimate the relative importance of anatomical noise. To obtain a more realistic estimate, 60 mammograms in CC projection were analyzed and $\sigma_{clutter,M}$ was calculated in a $60 \times 60$ mm$^2$ region for each image and a few values of $M$. At a ROI size of 1 mm ($M = 20$ since the pixel size is 50 $\mu$m) it was found that the variation expressed in glandular fraction $g$ was $\sigma_{g,M=20} = 0.06$. The standard deviation showed a linear relation in the range of $M$-values between 20 and 100, and extrapolation to M=1 gave $\sigma_{g,1} = 0.08$. To calculate $\sigma_{clutter,M}$, Eq. 4 is integrated over a gaussian distribution of $g$ with mean 0.5 and standard deviation $\sigma_{g,M}$.

Due to the different characteristics of quantum noise and anatomical noise, and that the two types of noise are affected very differently in DES, we expect to see a large dependence of our figure-of-merit on the ROI size.

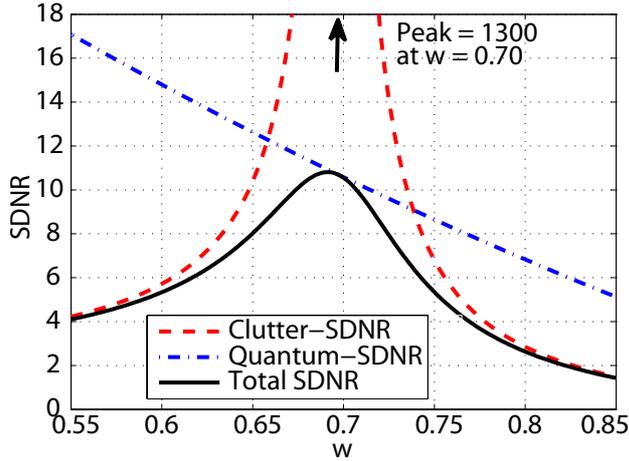 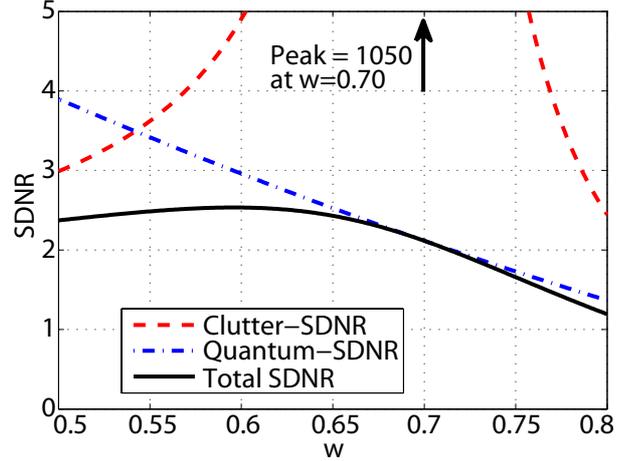

**Figure 3.** Calculated $SDNR_{tot}$ as a function of the weighting factor. In this case $M = 20$, i.e. 1 mm ROI size. Also included is SDNR taking into account only statistical and anatomical noise, respectively.

**Figure 4.** Calculated $SDNR_{tot}$ as a function of the weighting factor. In this case $M = 4$, i.e. 0.2 mm ROI size. Also included is SDNR taking into account only statistical and anatomical noise, respectively.

### 2.5. MPL filtering

A multi-prism lens (MPL) consists of two rows of prisms facing each other at an angle and symmetrically arranged around the optical axis (Fig. 2). The lens halves touch at the entrance side and are separated by a distance at the exit side of the lens.[12] Rays entering the lens at the periphery will encounter more prisms than will central ones, and will therefore experience a larger refraction. Imaging with an MPL is one dimensional, and since the refractive index varies as $E^{-2}$ at energies ($E$) and lens materials considered here, the lens is chromatic, thus having different focal lengths for different x-ray energies. By placing a slit at the image plane of a particular x-ray energy ($E_{peak}$) this energy can be picked out, whereas other energies are out of focus and preferentially blocked by the slit. The focal length, and thereby $E_{peak}$, can be tuned by varying the angle between the lens halves in an otherwise fixed geometry.

Due to practical and time constraints a DES image acquired using the MPL could not be prepared for this investigation. Nevertheless, Fig. 2 shows a schematic of an MPL set-up which is theoretically investigated in this study. A 61 mm long epoxy lens with teeth of height 100 $\mu$m is placed 759 mm from a 24.5 $\mu$m source, thus providing a 10 $\mu$m image 310 mm from the lens at a focal length of 220 mm according to the Gaussian lens formula. A slit of the same width as the image is placed in the image plane, and $E_{peak}$ is tuned by changing the angle between the lens halves. The spectrum of the filtering set-up is calculated using a geometrical model,[13] and is adjusted so as to center on the K-edge of iodine. Published linear absorption coefficients[16] and semi-empirical data on atomic scattering factors[17] served as input to calculate absorption and refraction. All MPL filtered spectra are additionally filtered with aluminum in order to reduce the low energy tails.

### 3. RESULTS

#### 3.1. Theoretical results

Our dual-energy figure-of-merit, $SDNR_{tot}$, was calculated for a number of different detector conditions and x-ray spectra. The same quantity was also calculated for the case of plain absorption imaging, and an SDNR-gain could be calculated. The results are summarized in Table 1 for an AGD of 1 mGy in all cases. The SDNR-gain is given for $M$-values 1 and 20, corresponding to the pixel size 50 $\mu$m and a ROI size of 1 mm, respectively. Note the vastly different SDNR-gains associated with these two scales. The dependence on scale is further illustrated in Figs. 3 and 4, where $SDNR_{tot}$ is plotted as a function of $w$ for ROI sizes of 1 mm and 0.2 mm, respectively. Also included is the SDNR taking only statistical noise into account as well as the clutter-only SDNR. In both cases the quantum noise is completely dominant. Since clutter is removed so effectively, the dependence on $w$ is not so large. In fact, for the 0.2 mm ROI, clutter noise is so relatively insignificant that the optimum $w$ is shifted from the value that minimizes the clutter residual.

Table 1. Calculated SDNR at an AGD of 0.5 mGy for different spectra and detector parameters..

| | $w_{opt}$ | **SDNR**$_{tot}$ **gain** | | **SDNR**$_{tot}$ rel. ideal |
|---|---|---|---|---|
| | | pixel | 1 mm ROI | 1 mm ROI |
| 45 kV, 2 mm Al, Ideal detector | 0.60 | 0.91 | 12.8 | 1.00 |
| 45 kV, 2 mm Al, Spectroscopic detector used for exp.[†] | 0.61 | 0.83 | 11.6 | 0.91 |
| 45 kV, 2 mm Al, Si, perfect E-resolution.[‡] | 0.70 | 0.53 | 7.3 | 0.55 |
| 45 kV, 2 mm Al, Si, no anti-coincidence | 0.78 | 0.39 | 5.4 | 0.41 |
| 45 kV, 2 mm Al, Si, no charge-sharing | 0.60 | 0.70 | 9.6 | 0.73 |
| 45 kV, 2 mm Al, Si (used in exp.) | 0.70 | 0.48 | 6.6 | 0.50 |
| 45 kV, 4 mm Al, Si | 0.71 | 0.56 | 7.6 | 0.60 |
| 40 kV, 4 mm Al, Si | 0.77 | 0.57 | 7.9 | 0.58 |
| 50 kV, 2 mm Al, Si | 0.66 | 0.43 | 5.9 | 0.45 |
| 50 kV, 4 mm Al, Si | 0.67 | 0.49 | 6.6 | 0.51 |
| 45 kV, lens + 1 mm Al, Si | 0.79 | 0.64 | 9.2 | 0.70 |
| 45 kV, lens + 1 mm Al, Si, perfect E-resolution[‡] | 0.79 | 0.79 | 10.8 | 0.83 |

[†] Non-pixelated CZT-detector with multi-channel analyzer
[‡] Perfect energy resolution, no threshold variation, but charge-sharing included

For conventional absorption imaging the opposite holds; anatomical clutter noise clearly dominates except for $M$ close to one. This means that in terms of noise our SDNR gain essentially boils down to $\sigma_{clutter}^{abs}/\sigma_{quant}^{DES}$, and that maximizing the iodine signal is more important than minimizing the clutter.

### 3.2. Measurements

#### 3.2.1. Iodine signal

The iodine signal was measured with a homogenous 50% distribution of PMMA in the box in front of the iodine containers. Fig. 5 shows the absorption contrast as a function of depth of the iodine containers for both detectors, along with the model predictions for an iodine concentration of 3 mg/ml. If the model concentrations are matched to the measured signals, concentrations of 3.45 and 3.75 mg/ml are obtained for the CZT and silicon detectors respectively. These concentrations are reasonable considering the measurement uncertainty when preparing the iodine solution. The DES signals are shown in Fig. 6. A gaussian energy resolution with a full-width-at-half-maximum (FWHM) of 1 keV for both detectors, and a threshold resolution of 0.8 keV FWHM matched the model to the least square fit of the experimental results when assuming the above iodine concentrations. Since the silicon detector does not suffer from hole tailing, a somewhat better energy resolution can be expected than for the CZT detector. The threshold resolution is on the other hand most likely worse than 0.8 keV.

Variations in the detected signals are likely due to two different effects; variation in the x-ray tube output and, for the silicon detector, drifting energy thresholds. Due to a limited detector area and count rate, in particular for the CZT detector, the measurements required several hours each, and the tube output cannot be assumed constant during this time. A drifting low-energy threshold of the silicon detector would affect the total number of detected photons, thus altering absorption and subtraction signals. A drifting high-energy threshold affects the balance between the high- and low-energy bins, and thus the subtraction signal.

#### 3.2.2. Wedge phantom

Measurements on the PMMA-to-oil wedge phantom isolated $\sigma_w$ so that $w_{\text{opt}}$ could be determined and the effect of anatomical noise subtraction visualized. Fig. 7 shows the standard deviation of the wedge signal for PMMA fractions 0.1-0.9 as a function of $w$. Optimal weight factors are found at the minima and are $w_\text{opt} = 0.63$ and 0.72 from experiment for the CZT and silicon detectors respectively. The effect of the weight factor is illustrated in Fig. 8, showing $\Delta S_\text{s}$ between the wedge and a PMMA fraction of 0.5 for weight factors differing from the optimal one.

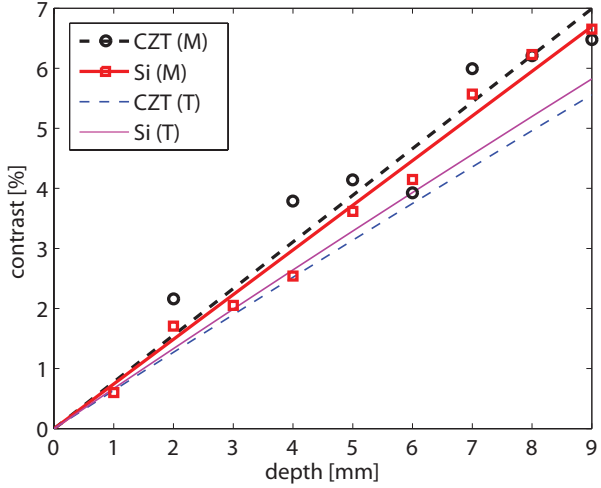
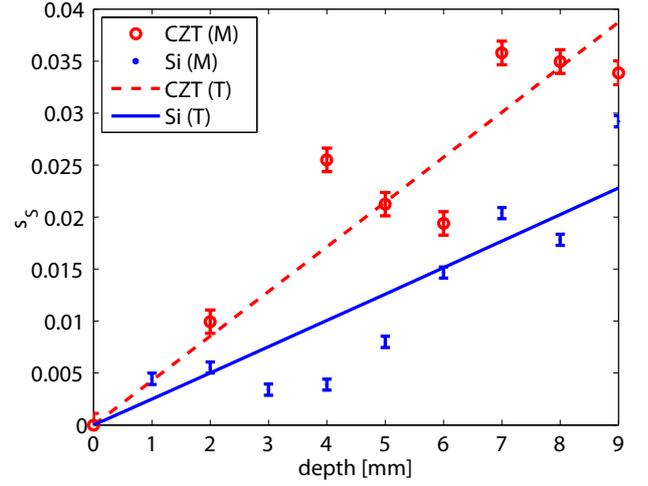

**Figure 5.** The absorption contrast as a function of depth of the iodine containers for both detectors, along with model predictions for an iodine concentration of 3 mg/ml. Outliers were removed from the data set. Statistical errors are too small to be visualized with error bars.

**Figure 6.** The DES signal difference ($\Delta S_\text{S}$) as a function of depth of the iodine containers. The model has been matched to fitted experimental results. Error bars correspond to one standard deviation statistical uncertainty. The DES signals of the two detectors refer to slightly different iodine concentrations and are therefore not directly comparable.

### 3.2.3. Anatomical clutter phantom

To simulate anatomical clutter, the PMMA box was filled with 5 and 10 mm PMMA cylinders which were immersed in olive oil. The resulting absorption image at an AGD of 0.5 mGy is shown in Fig. 9, and Fig. 10 shows the DES image calculated with $w = 0.72$ as determined from the wedge phantom. The 5 deepest iodine containers (down to 5 mm) are clearly visible. Containers 4 mm and below might also be visible, but in that case several false positives with the same contrast are present in the image. These false positives are again due to drifting energy thresholds and a changed tube output. In fact, the scan stripes of the detector are clearly visible in the image. Although severe in this study, these problems would have less impact in a full-scale system where the signals from many detectors are added and acquisition times are in the order of 10 s.

To compare the phantom structure to the theoretical clutter noise based on real mammograms, a 400 by 400 pixel region without iodine containers were used to calculate the PMMA-fraction standard deviation $\sigma_{g,M=20}$. The result was 0.08, which is close to the value of 0.06 from real mammograms. Although a crude phantom, this shows that the resemblance to a real anatomical background is good enough for the present study.

### 3.2.4. Dual energy figure of merit

From the clutter image, a $20 \times 20$ mm$^2$ region without iodine containers was selected for calculation of $\sigma_{tot,M}$ for a number of $M$-values in the range 1 to 40. Fig. 11 shows $\sigma_{tot,20}$ as a function of the weight factor (note that this measurement is different from the one in Fig. 7 although the plots look similar). The quantum noise seems to be about twice as high as expected from theory, which was corroborated by a similar measurement in a DES image without clutter. This is due to correlated detector noise which was not included in the theory. The detector noise can be incorporated into $\sigma_{clutter,M}$ by substituting $M$ for an effective $M_\text{eff}$ in Eq. 7. For $M =[2, 4, 10, 20]$ we measured $M_\text{eff,DES} =[1.9, 3.5, 6.2, 7.8]$ in a DES image without clutter. For the conventional absorption image of the same phantom we got $M_{\text{eff,abs}} =$ [2.0, 4.0, 9.5, 17], which shows that this effect is mainly in the DES image.

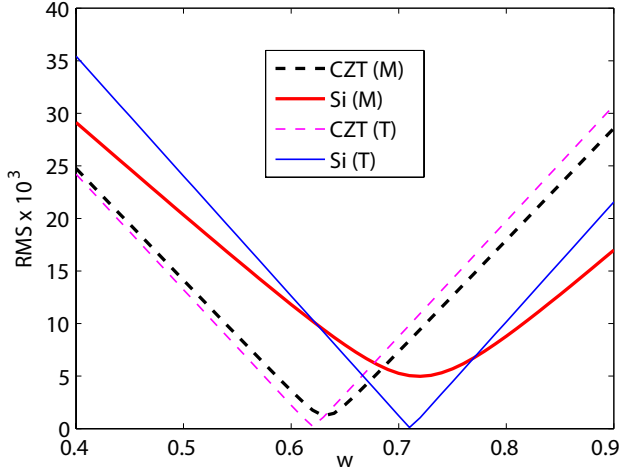

**Figure 7.** The standard deviation of the DES signal from the PMMA-to-oil wedge phantom as a function of the weight factor ($w$) for both detectors as determined by experiment and modeling.

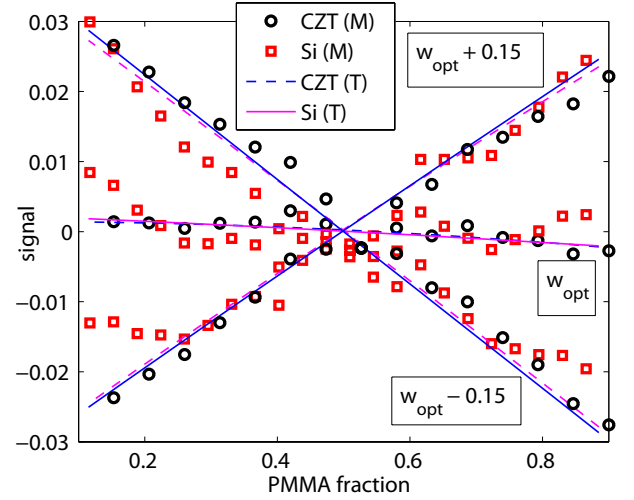

**Figure 8.** The signal difference between a range of PMMA fractions and a fraction of 0.5 for the optimal weight factor and factors differing from the optimum.

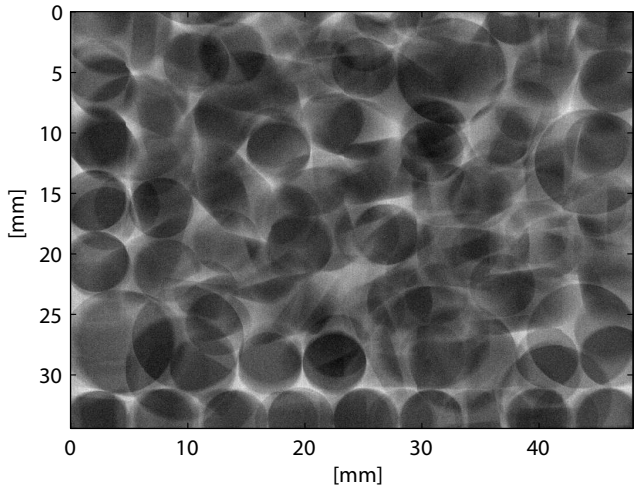

**Figure 9.** An absorption image of the anatomical clutter phantom with 1-9 mm iodine containers at 45 kV and 2 mm aluminum filtration.

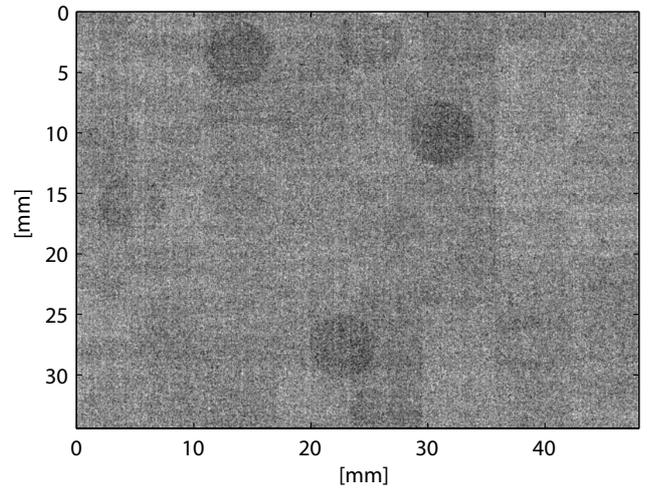

**Figure 10.** The corresponding DES image calculated with $w = 0.72$. The five deepest holes are visible.

The measured total noise and the iodine DES signal difference gives the SDNR$_{tot}$. The same measurement in the plain absorption image enables a calculation of the SDNR gain, which is plotted as a function of $M$ in Fig. 12. Also included are the teoretical SDNR gain curves, with and without correction for the measured $M_{\text{eff}}$. As can be sen in the image, $M_{\text{eff,DES}}$ overcorrects the quantum noise. We believe this is due to a particulary high upper threshold variation in the image used for the measurement of $M_{\text{eff,DES}}$.

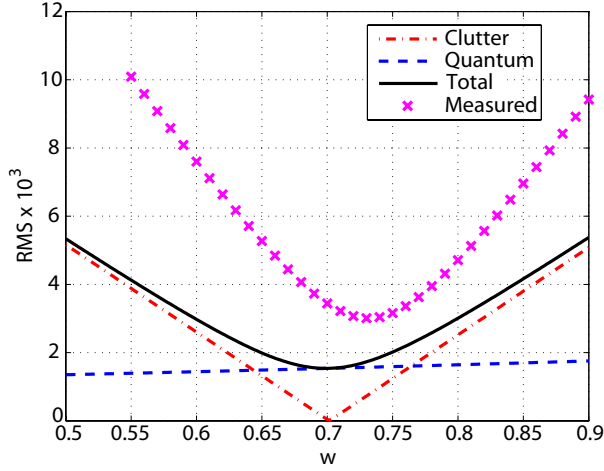 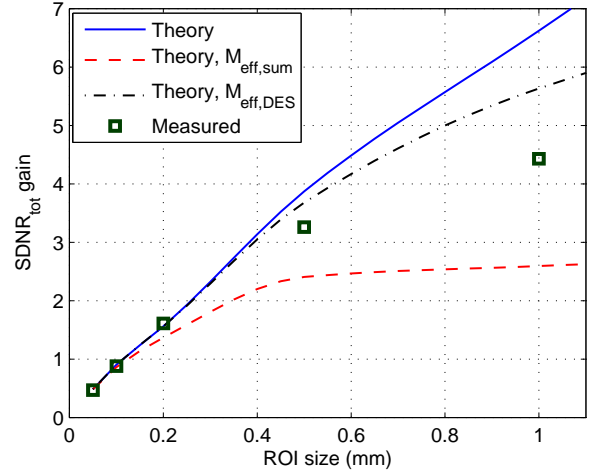

**Figure 11.** Total noise measured in the clutter DES image as a function of the weighting parameter, compared to the theoretical noise components and sum. ROI size is 1 mm ($M = 20$).

**Figure 12.** Measured SDNR gain of the DES image compared to the absorption image as a function of ROI size. Also included are theoretical values, with and without compensation for the measured $M_{\text{eff}}$.

### 3.3. Lens results

An MPL filtered 45 kV tungsten spectrum with 1 mm Al additional filtration is shown in Fig. 13 together with spectra filtered with 2 and 4 mm of aluminum. All spectra in Fig. 13 are normalized to unity total intensity, to visualize the spectral shape. The relative efficiency of the filtering methods can be appreciated from Fig. 14, which indicates that the lens increases the flux approximately an order of magnitude at the gain peak due to its focusing properties. This, however, is not necessarily the case since the lens spectrum was measured using a smaller focal spot, which is essential to obtain a narrow gain peak.

The theoretical efficiency of the MPL filter for DES imaging is summarized in Table 1. If the lens is coupled to the silicon-strip detector it increases the SDNR by 40% compared to the 45 kV spectrum filtered with 2 mm aluminum. The limited energy resolution reduces the SDNR by 15% compared to 10% for the conventional spectrum. This is to be expected since a limited energy resolution is more important for a spectrum concentrated around the K-edge.

### 4. DISCUSSION

Two detectors were used in this study; a non-pixelated CZT detector to simulate perfect conditions for ES, and a silicon detector for simulating realistic conditions for medical imaging. For the latter, a lower quantum efficiency, a somewhat worse energy resolution, and charge sharing combined degrade the SDNR by 45%. Charge-sharing alone is responsible for 30% SDNR loss, but the effect is mitigated by the anti-coincidence logic of the ASIC. Without this feature each shared event would be registered as either two lower-energy photons or one in each energy bin, and the SDNR loss due to charge sharing would be 45%. An improvement of the logic with respect to DES imaging would be to discard all shared photons, to put them in a separate bin, or, ideally, to add the energy of the two photons. In the present implementation, the merit of the anti-coincidence logic is primarily to improve spatial resolution and provide the correct statistical weight to each photon.

Since the difference in linear attenuation coefficients between adipose and fibroglandular tissue is similar to the difference between PMMA and olive oil, in the ideal case tissue subtraction as well as iodine DES signal would be similar to a real breast. This is not totally correct since the iodine solution contains water and when using a weight factor optimized to reduce the contrast between PMMA and olive oil, the water DES signal will be negative relative the iodine signal The iodine signal is thus reduced by approximately a factor of two in the present study. In the real case, this will be of less importance since the linear attenuation of both water and cancerous tissue is similar to the attenuation of glandular tissue and will thus yield a DES signal much smaller than the one from iodine.

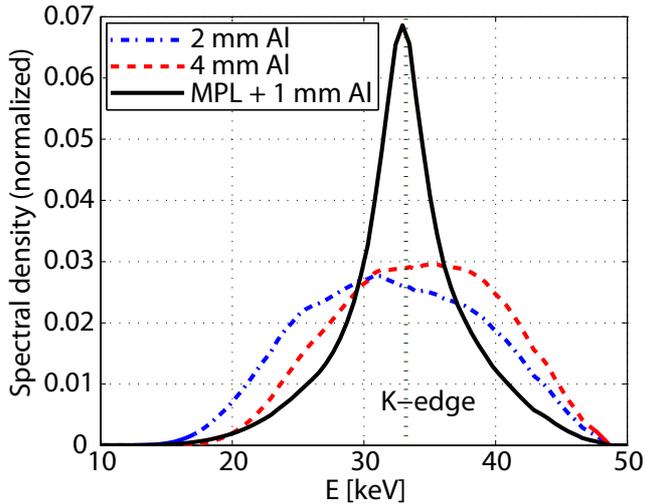 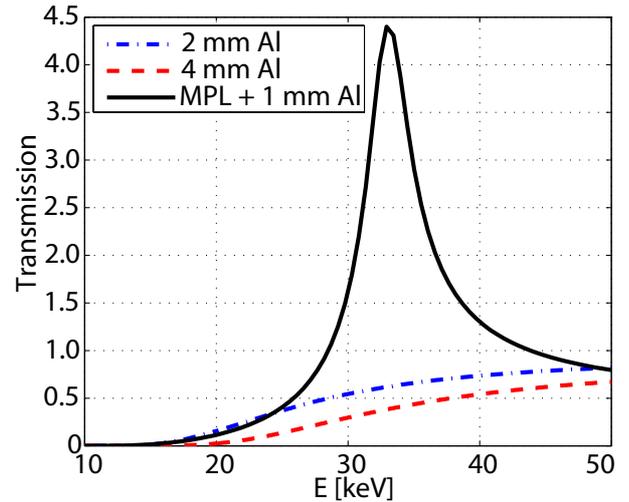

**Figure 13.** An MPL + 1 mm aluminum filtered 45 kV tungsten spectrum together with absorption filtered spectra. The integrated density of all spectra is normalized to unity.

**Figure 14.** The transmission of the MPL compared to absorption filtering.

Generally speaking, a narrower spectrum that is centered on the absorption edge is more efficient for DES imaging than a broader one. The MPL filter therefore improves the SDNR compared to realistic absorption filtered spectra (Table 1). One should note, however, that a real detector with imperfect energy resolution cannot make much use of the energy region near the K-edge. Thus, there is a tradeoff since this region will constitute a larger fraction of the spectrum if the spectrum is narrowed down. This may indicate that the lens would fair better if we use two spectra peaking some distance away from the K-edge on opposite sides in a DS approach.

It is not the objective of this work to compare the electronic spectrum splitting method to the dual spectra method. Nevertheless, it can be argued that a drawback of the ES method is that the dose cannot be distributed freely between the low and high energy regions in order to optimize the SNR, such as is the case for the DS method. This, however is not a serious drawback. If one scales the part of the spectrum above the K-edge by a constant factor, while keeping the total AGD fixed, one will find that it is optimal to double the dose in the high energy region. This, however, will only lead to a SDNR gain of 8%.

## 5. CONCLUSIONS

Contrast-enhanced dual-energy subtraction imaging using electronic spectrum splitting has been investigated in a mammography model with an iodinated contrast agent. A clinically feasible silicon detector was compared to a near-ideal one, and the improvement over absorption imaging was estimated using a signal-difference-to-noise ratio that takes into account statistical as well as anatomical noise. A theoretical model of the detectors was benchmarked to the measurements, and the model was used to investigate the influence of different detector effects and the potential improvement if introducing an energy filter based on a multi-prism x-ray lens instead of plain absorption imaging.

The silicon detector provides about 50% SDNR compared to the ideal case. Although far from ideal, it was seen in the study that the detector still improves the detectability of iodine on a lumpy background. An MPL filter coupled to the silicon detector would improve the SDNR 40% compared to the spectrum used in the experiment. Although a much larger improvement is possible with perfect energy resolution, this is still significant considering the fact that the experimental spectrum already has substantial absorption filtering with a correspondingly low dose rate. A thorough investigation that includes the different geometrical constraints on the MPL setup is, however, necessary before any final conclusions can be drawn on this topic. The limited energy resolution of the detector reduces the possible improvement by the MPL filter, and it might be more advantageous to employ the filter in a DS approach. This is, however, left for future studies.

## ACKNOWLEDGMENTS

The authors wish to thank Alexander Chuntonov for setting up the silicon-strip detector.